\documentclass[twocolumn,showpacs,amsmath,amssymb]{revtex4}
\usepackage{graphicx}
\usepackage{epsfig}
\usepackage{subfigure}
\usepackage{flafter}
\usepackage{dcolumn}
\usepackage{float}
\usepackage{mathrsfs}
\begin{document}

\title {Orientational correlations and the effect of spatial gradients in the equilibrium steady state of hard rods 
in 2D : A study using deposition-evaporation kinetics}

\author{Mahendra D. Khandkar and Mustansir Barma}
 \affiliation{
 Department of Theoretical Physics, Tata Institute of Fundamental Research, Homi Bhabha Road, Colaba, Mumbai-400 005, INDIA.}

\date{\today}

\begin{abstract}

 Deposition and evaporation of infinitely thin hard rods (needles) is studied 
in two dimensions using Monte Carlo simulations. The ratio of deposition to
evaporation rates controls the equilibrium density of rods, and
increasing it leads to an entropy-driven transition to a nematic phase in which
both static and dynamical orientational correlation functions decay as power laws, with
exponents varying continuously with deposition-evaporation rate ratio. Our results for
the onset of the power-law phase agree with those for a conserved number of rods. At a 
coarse-grained level, the dynamics of the non-conserved angle field is described
by the Edwards-Wilkinson equation. Predicted relations between the 
exponents of the quadrupolar and octupolar correlation functions are
borne out by our numerical results. We explore the effects of spatial inhomogeneity
in the deposition-evaporation ratio by simulations, entropy-based arguments and a study of the new terms
 introduced in the 
free energy. The primary effect is that needles tend to align along the local spatial gradient of the ratio.
A uniform gradient thus induces a uniformly aligned state, as does a gradient which varies randomly
in magnitude and sign, but acts only in one direction. Random variations
of deposition-evaporation rates in both directions induce frustration, resulting in a state with
 glassy characteristics.

\end{abstract}


\maketitle

\section{\label{sec:level1} Introduction }

 An assembly of particles interacting via hard-core repulsion serves as a useful model for studying
simple fluids, colloids, liquid crystals and many other soft matter systems.
The analysis of such model systems helps in understanding the features of real systems,
such as their phase behaviour, structural and  dynamic properties.
 An important role is played by the anisotropy of shape of the constituent particles, which 
 can range from thick elongated platelets to thin 
rods. Some examples of systems in which the constituent particles show anisotropy are certain types of
 colloids, liquid crystals and
 protein molecules.  In particular, rod-like particles are found in suspensions of the
tobacco mosaic virus~\cite{fraden}, nematic liquid crystals~\cite{gennes} and, recently, carbon nanotube gels~\cite{islam}.
 All these systems show very rich and characteristic phase behaviour.

 Rod-like particles have been modeled theoretically as ellipses~\cite{cuest}, rectangles and spherocylinders
~\cite{stroo,bates,lagom} with varying aspect ratios, a limiting case being infinitely thin hard
 rods or needles~\cite{eppen}. These systems exhibit a number of interesting entropy-driven
 phase transitions  which have been studied in two and three dimensions, usually using simulations with 
number-conserving dynamics. On the other hand, there are a number of physical processes which involve
adsorption (deposition) and desorption (evaporation) of particles, which do not conserve 
particle number and which are important for some monolayer growth processes. Adsorption and desorption
are also important in the binding/unbinding of ligands to microtubules, the interaction of 
proteins with DNA~\cite{evans,efrey} and many catalytic reactions. 
Finally, in recent experiments on assemblies of long objects (rice grains, thin metal rods)
on a vibrating plate~\cite{swamy}, individual particles jump off and return to the plate, leading 
ultimately to a state with interesting patterns.
These considerations motivate us to study
the deposition and evaporation of hard objects with rigid boundaries on a substrate. While a deposition-only system, of the
type studied in random sequential adsorption~\cite{evans}, can end up in a non-evolving jammed configuration, with the addition
of evaporation, the system eventually reaches an equilibrium steady state with a density governed
by the rates of deposition and evaporation [12-17]. While most of these studies have focussed on the kinetics of 
approach to steady state, in this paper, we are interested in the properties of the steady state itself.
 Specifically, we study the patterns formed due to deposition and evaporation 
of infinitely thin hard rods (needles) on a 2D substrate. Needles are a limiting case of 
rod like particles in the systems mentioned earlier. Though not directly applicable to any physical system, this 
is an important limiting case; the limit of zero width simplifies the problem by 
eliminating the aspect ratio as a parameter.
The hard core constraint is enforced by rejecting any deposition event which results in an overlap of needles.

It is useful to recall some known facts about a system of hard needles with no externally imposed spatial
inhomogeneities. This system shows
a transition from a low-density orientationally disordered (isotropic) state to a high-density
ordered state with nematic correlations. This transition, whose existence was pointed out by
Onsager~\cite{onsag}, can be viewed as an outcome of the interplay between orientational and translational entropy
 of the needles; the ordered (aligned) state is preferred at high density since alignment leads to an increase of
translational entropy, albeit at the cost of orientational entropy. The
nature of the orientational ordering is dimension-dependent. In three dimensions, orientational
long range order (LRO) sets in. A state with LRO would break the continuous symmetry of
rotations, and is thus not expected to occur in 2D, even though the Mermin-Wagner theorem cannot be
generalized to this system~\cite{stral}. 
Indeed, the simulation study of Frenkel and Eppenga~\cite{eppen} on a system with a fixed number of needles 
confirms the absence of LRO in 2D, and
 finds a phase with power-law decays of orientational correlations, quite analogously to  
the XY model~\cite{nelso}.     

On a coarse-grained scale, the local orientation at location {\bf r} and time {\it t} is specified
by an angle field $\theta({\bf r},t)$. 
The orientational correlation functions of interest are defined as 
\begin{equation}\label{eq:eq1}
 g_\ell({\bf r},t) = \langle cos[\ell(\theta({\bf r},t) - \theta(0,0))]\rangle
\end{equation}

\noindent where $\ell$ is an even integer, and $\theta$ and $\theta + \pi$ represent the same state. 
Quadrupolar correlations are probed by $\ell$ = 2, whereas higher values of $\ell$ correspond
to higher multipolarities.
From numerical simulations, we find power law decays in both space and time: 
$g_\ell(r,0) \sim r^{-\eta_\ell}$ and $ g_\ell(0,t) \sim t^{-\beta_\ell}$ for both $ \ell = 2 $ and 4.
Our results for the static correlations conform to the Kosterlitz-Thouless theory for the onset
of correlations, while our results for the dynamics show that their decay in time is governed by the
Edwards-Wilkinson equation.                                                                                                                             
We also study spatial variations of the deposition rate, and find strong effects on the nature of the ordering.
 We consider several types of variations:
(i) a sharp change across a linear interface ; (ii) a smooth linear gradient; (iii) a random variation 
of rates in one direction and (iv) random variation of rates in the plane. We find that the qualitative effect
of spatial variations is to induce alignment of needles in the direction of gradient, an effect with an
entropic origin. In (i), the effect dies down slowly with increasing distance from the interface, but in cases (ii) 
and (iii) it results in a state with overall alignment along the direction of variation of the deposition rate.
 The random variation in (iv) induces frustration and 
the result is then a state with glassy features such as strong initial condition dependence and slow relaxation.
 
\section{\label{sec:level1} MODEL AND PROCEDURE }

In our model, infinitely thin hard rods (needles) are added to a 2D substrate with area $A$ with
a constant attempt rate, and simultaneously removed randomly from the substrate with a specified rate.
In a deposition attempt, the location of the centre of mass of the needle is chosen at random on the
substrate, and the orientation angle is chosen at random as well. Let $\Gamma_{d}$ be the rate 
of attempted depositions per unit area per unit angle interval.  An attempt is successful only if 
the depositing needle in question would not overlap with existing needles on the substrate; otherwise it is
rejected.

During evaporation, a needle is chosen at random from those present on the substrate and then removed. 
If the total rate of such removals is $R_{e}$ we may associate a removal rate $\Gamma_{e}$ =$ R_{e} \over 2\pi A$ 
per unit area per unit angle interval. The ratio

\begin{equation}\label{eq:eqkd}
 {\kappa} = {\Gamma_{d} \over \Gamma_{e}} 
\end{equation}

\noindent of deposition to evaporation rates is the control parameter in the problem.
As we show below, $\kappa$ is related to the fugacity $z$ = $e^{\beta \mu}$ of an equilibrium grand canonical
system.

The model under consideration can be thought of as describing the adsorption and release of needle-like 
gas molecules on a substrate in contact
with a gas reservoir with which it can exchange particles. The equilibrium state on the substrate is
then described by the grand canonical ($\mu A T$) ensemble, where $\mu$ , $A$ and $T$ are respectively the
chemical potential, substrate area and temperature. Define scaled coordinates ${\bf s}_{i}$ = 
${\bf r}_{i} /L$, where ${\bf r}_{i}$ is the position of the $i^{th}$ particle on the substrate and 
$L$ is the linear dimension of the system. 
The grand canonical partition function can be written~\cite{landau} as 

\begin{eqnarray}
{\mathscr{Z}} = {\sum_{N=0}^{\infty}} {z^N \over N!} {(A/\Lambda^2)^N} {\int d{\bf s}_1 \int d{\bf s}_2 ....
\int d{\bf s}_N}
 \nonumber \\
{\int d\theta_1 \int d\theta_2 .....\int d\theta_N} {e^{-\beta U({\bf s}_1,{\bf s}_2,...{\bf s}_N,\theta_1,
\theta_2,....\theta_N)}}
\end{eqnarray}

\noindent where $\Lambda$ = ($2 \pi \hbar^2 \over m k_B T$$)^{1/2}$ is the thermal wavelength which results
from integrating over the momentum of needles,
and $U$ is the interaction energy for a configuration in which there are $N$ needles with scaled 
centre of mass locations $({\bf s}_1,{\bf s}_2,...{\bf s}_N)$ and orientations $(\theta_1,\theta_2,....\theta_N)$.
The corresponding equilibrium probability density of a configuration $C \equiv ({\bf s}_1,{\bf s}_2,...{\bf s}_N,
\theta_1,\theta_2,....\theta_N)$ is~\cite{landau} 

\begin{equation}\label{eq:eqN}
{P_{eq}(C)} = {1 \over \mathscr {Z}}  {z^N \over N!} {(A/\Lambda^2)^N}{e^{-\beta U({\bf s}_1,{\bf s}_2,...{\bf s}_N,
\theta_1,\theta_2,....\theta_N)}}
\end{equation}

\noindent For our system of hard core needles, the interaction energy $U$ $\rightarrow$ $\infty$ when needles
overlap, while $U = 0$ when there is no overlap between any needles. Thus all allowed configurations with
fixed $N$ have equal weights.

 Deposition - evaporation dynamics involves changes of $N$.
The evolution from a configuration $C$ to a configuration $C'$ can be described by the master equation 

\begin{equation}\label{eq:eqM}
{\partial P(C) \over \partial t} = {\sum_{C'}} {[P(C') W(C' \rightarrow C) - P(C) W(C \rightarrow C')]}
\end{equation}

\noindent The steady state of Eq. (\ref{eq:eqM}), obtained by setting $\partial P(C) \over \partial t$ = 0, 
is in fact an equilibrium state if the condition of detailed balance 
\begin{equation}\label{eq:eqd}
{W(C' \rightarrow C) \over W(C \rightarrow C')} = {P_{eq}(C) \over P_{eq}(C')} 
\end{equation}

\noindent is satisfied for every pair of configurations $C$ and $C'$ that can be reached from each other.
 Now, let $C$ denote an $N$-needle  configuration and let $C'$ be the $(N-1)$-needle configuration obtained
from $C$ by removing a particular needle. Using Eq.(\ref{eq:eqN}) in Eq.(\ref{eq:eqd}), we see that 
$\Gamma_{d} \over \Gamma_{e}$ = $z A \over \Lambda^{2} N$.
 Thus, the steady state of deposition-evaporation dynamics is described by the 
grand canonical equilibrium state with 
\begin{equation}\label{eq:eqK}
 {\kappa} = {z \over \rho\Lambda^{2}} 
\end{equation}

\noindent where $\rho$ = $N \over A$ is the areal density of needles.

In our Monte Carlo studies we varied the control parameter $\kappa$ in the range 1 to 40 and 
monitored the resulting density and orientational correlations.
We used an $L \times L$ substrate (with $L$ = 15 and 25) where $L$ is in units of needle length. 
For each value of $\kappa$ we made multiple runs, 
allowing up to $10^{7}$ Monte Carlo time steps for equilibration. The Monte Carlo time $t$ is defined as the number of
attempts divided by $L^{2}$. Averaging was done over 10 sets of 
independent runs and 1000 configurations from each run after equilibration. 

\begin{figure}[htbp]
\subfigure[]{
\label{fig:subfig:rho_k_a}
\includegraphics[height=5cm, width=7.2cm]{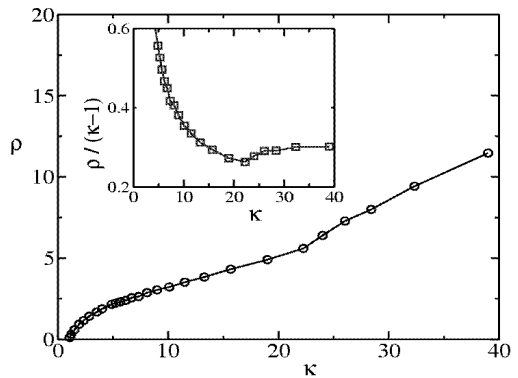}}
\subfigure[]{
\label{fig:subfig:rho_k_b}
\includegraphics[height=5cm, width=7.2cm]{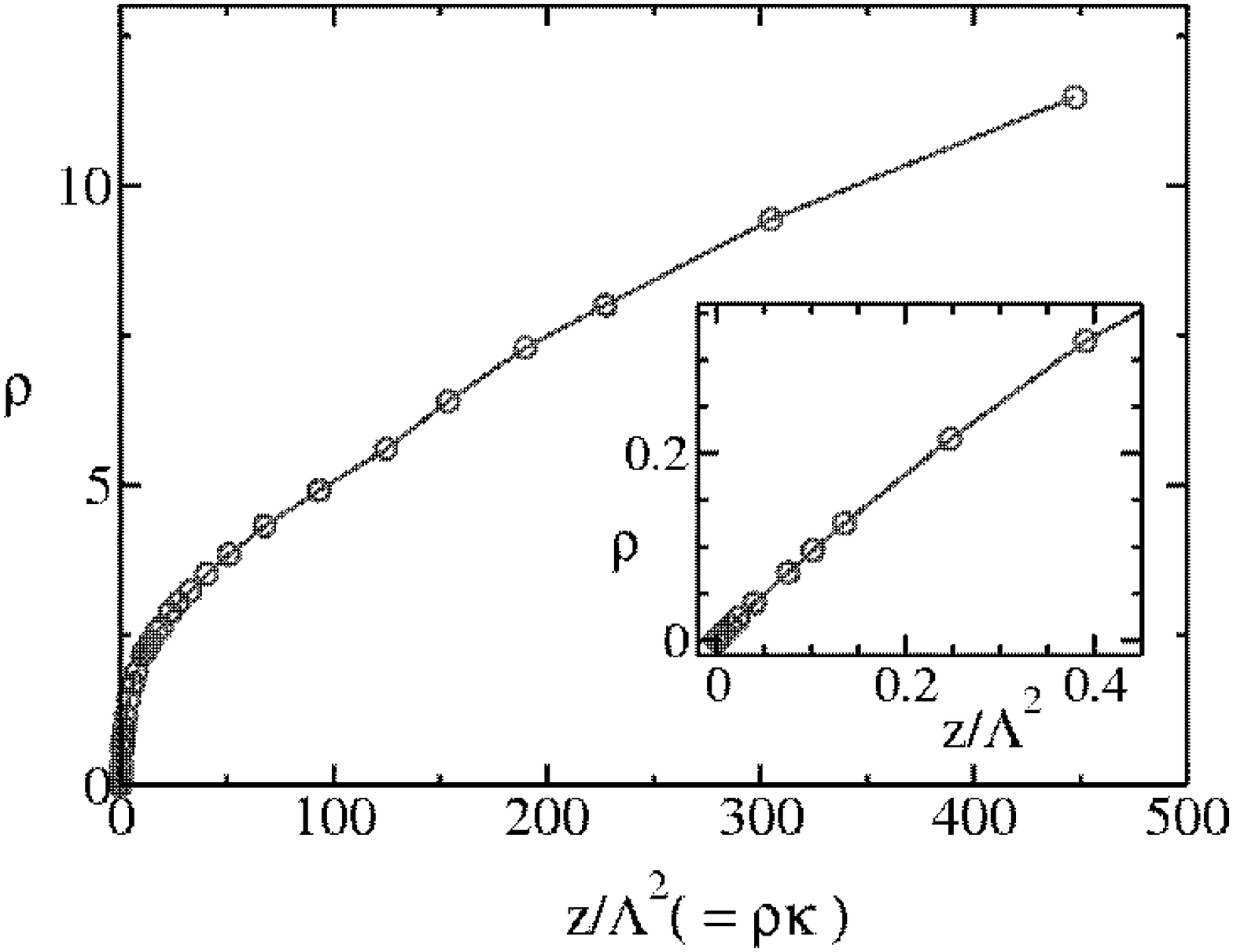}}
\caption{ \small (a) The variation of $\rho$ (number of rods per unit area) with $\kappa$ shows a change of behaviour for $\kappa$ in the range 20-25.
 This is more prominently depicted in the inset which shows the variation of $\rho$/$(\kappa-1)$.
 (b) Variation of $\rho$ with $z/\Lambda^2$. The inset shows the initial portion of the curve.} 
\label{fig:rho_k}
\end{figure}

\begin{figure}[htbp]
\subfigure[$\kappa \simeq $ 19]{
\label{fig:subfig:2a}
\includegraphics[height=4.2cm, width=4.2cm]{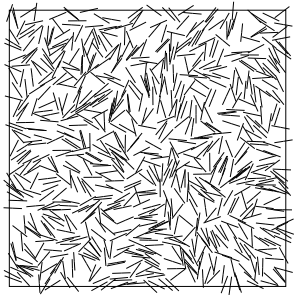}}
\subfigure[$\kappa \simeq $ 24]{
\label{fig:subfig:2b}
\includegraphics[height=4.2cm, width=4.2cm]{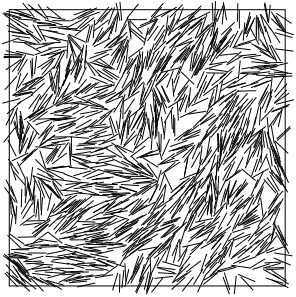}}
\subfigure[$\kappa \simeq $ 32]{
\label{fig:subfig:2c}
\includegraphics[height=4.2cm, width=4.2cm]{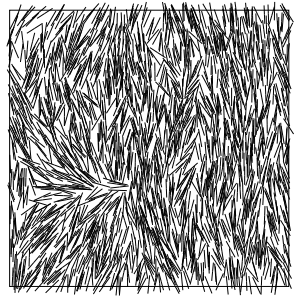}}
\subfigure[$\kappa \simeq $ 39]{
\label{fig:subfig:2d}
\includegraphics[height=4.2cm, width=4.2cm]{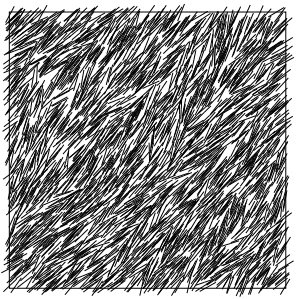}}
\caption {\small Snapshots of hard rod configurations at different values of $\kappa$. The substrate size
is 15$\times$15. Observe the formation of defects in the configurations (a), (b) and (c).}
\label{fig:confs}
\end{figure}

Figure \ref{fig:subfig:rho_k_a} shows the variations of the density with $\kappa$, while Fig. \ref{fig:subfig:rho_k_b}  shows $\rho$ 
plotted against $z/\Lambda^2 = \rho \kappa$. The inset in Fig. \ref{fig:subfig:rho_k_a} shows a marked change in the dependence
of $\rho / (\kappa-1)$ on $\kappa$ for $\kappa$ in the range 20-25. 
 As we shall see, there is a transition to a phase
with power law decay of orientational correlations beyond $\kappa=\kappa_{c}\simeq 25$, as illustrated by the
representative configurations shown in Fig. \ref{fig:confs} for different values of $\kappa$.  
We turn to a quantitative study of orientational ordering in the next section. 

\section{\label{sec:level1} ORIENTATIONAL ORDERING}

\subsection{\label{sec:level2} Order parameter} 

     For a system of N hard rods in 2D, the nematic order parameter $q$ is given by 
     
\begin{equation}\label{eq:eqq}
 {q} =  {1 \over N}  \langle {\sum_{i=1}^N} cos(2\theta_{i}) \rangle      
\end{equation}

\noindent where $\theta_{i}$ is the angle made by the ${\it i}^{th}$ rod with the nematic director, which
itself has an orientation $\phi$ with respect to a fixed reference X-axis.
 Both $\phi$ and $q$ can be found by studying the tensor order parameter defined as  

\begin{equation}\label{eq:eqQ}
{{\bf Q}_{\alpha\beta}}  = {1 \over N} \langle {\sum_{i=1}^N} [2u_{\alpha}(i)u_{\beta}(i)-
{\delta_{\alpha\beta}}] \rangle                                       
\end{equation}
                                                        
\noindent where $u_{\alpha}$(i) is the $\alpha^{th}$ component of ${\bf u}(i)$, the vector
specifying the orientation
of the ${\it i}^{th}$ rod. The eigenvalues of ${\bf Q}_{\alpha \beta}$ are $\pm q$, and the corresponding
eigenvectors pick out directions along and perpendicular to the director orientation $\phi$.
Insofar as there is no long range order in the 2D needle system, $q$
vanishes in the thermodynamic limit. In simulations on finite systems, though, $q$ may appear
to be non-zero (Fig. \ref{fig:subfig:2d}), over short times. Tracking the onset of such an apparent value is not  
a reliable way to locate the transition point.

\subsection{\label{sec:level2} Orientational cumulant of $q$ }

A better indication of the transition point, and also the nature of the ordered phase, is 
provided by monitoring the probability distribution functions $P_{L}(q)$ of $q$,
 where $q$ is the block-averaged
value of the local order parameter with the system divided into blocks of finite size $L$~\cite{binde3}. 
  A measure of the non-Gaussian
character of $P_{L}(q)$ is provided by the reduced $4^{th}$ order Binder cumulant of $q$ 
\begin{equation}\label{eq:eqU}
 {U_{L}} = 1 - { \langle q^{4} \rangle_{L} \over 3 \langle q^{2} \rangle^{2}_{L}}
\end{equation}

\noindent where $L$ is the linear size of the sub-system (block).  
$U_{L}$ provides a useful diagnostic tool to monitor the ordering induced by varying $\kappa$~\cite{ising}.

For $L \ll \xi$, $U_{L}$ is expected to stay close to a fixed point value $U^{*}$. 
Thus, the occurrence of a critical point with $\xi = \infty$ can be identified by plotting $U_{L}$ 
against $\kappa$ for various values of $L$, and looking  for common intersection point.

\begin{figure}[htbp]
\includegraphics[height=5cm, width=7cm]{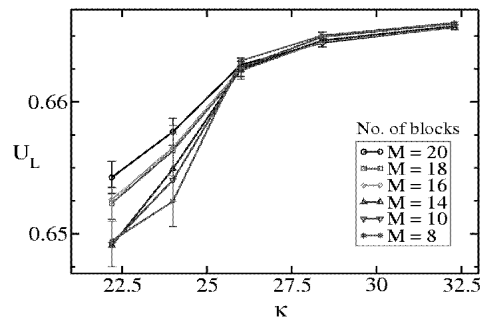}
\caption{ \small Orientational cumulants $U_{L}$ as a function of deposition-evaporation ratio $\kappa$
    for various subsystem sizes L = K/M. Note the collapse of the curves beyond $\kappa_{c} \simeq $ 25,
   pointing to the occurrence of a power law phase. }
\label{fig:cumulant}
\end{figure}

We analysed the Monte Carlo data of our model by 
monitoring $U_{L}$ (Eq. \ref{eq:eqU}).  The analysis was performed as follows :
 We simulated a single large system of size K $\times$ K (K = 25) and divided it into subsystems of
 size L $\times $ L,
 thereby having total $M^{2}$ number of subsystems with $M$ = K/L~\cite{weber}. Then, $M$ was incremented in integer steps
 starting from 1 and $U_{L}$ was estimated for those subsystem sizes L  where a good analysis is possible. Consequently,
we did not consider very small or very large values of $M$. Also, the curve for $M = 12$ lies 
anomalously low and was not included. 
 The number of subsystems ($M$) we use for estimating the cumulant range from 8 to 20.
 Fig. \ref{fig:cumulant} shows the variation of $U_{L}$ as a function of $\kappa$ for various values of L.

Below $\kappa_{c} \simeq 25.8$, the curves are separate and distinct, but they collapse at $\kappa_{c}$,
indicating the onset of ordering (Fig. \ref{fig:cumulant}). Moreover, the curves seem to stay collapsed for $\kappa > \kappa_{c}$
suggesting that $\xi$ remains infinite in this phase, i.e. this is a phase with power law decay of
correlations. Corroboration of this is provided by directly monitoring the correlation functions as 
described below.
 
\subsection{\label{sec:level2} Orientational correlation functions }

\begin{figure*}[htbp]
\subfigure[]{
\label{fig:subfig:g2r}
\includegraphics[height=5cm, width=7cm]{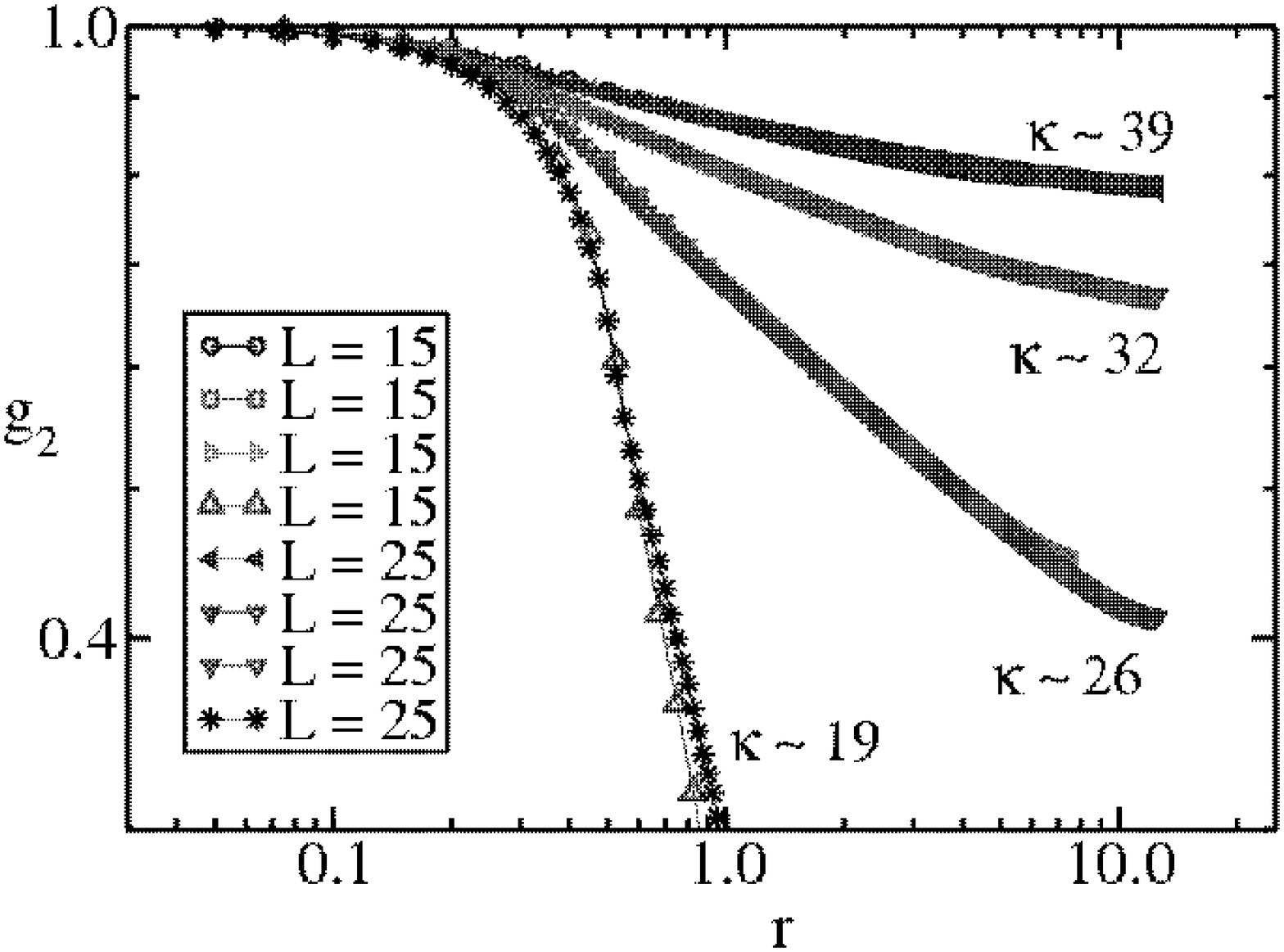}}
\subfigure[]{
\label{fig:subfig:g2t}
\includegraphics[height=5cm, width=7cm]{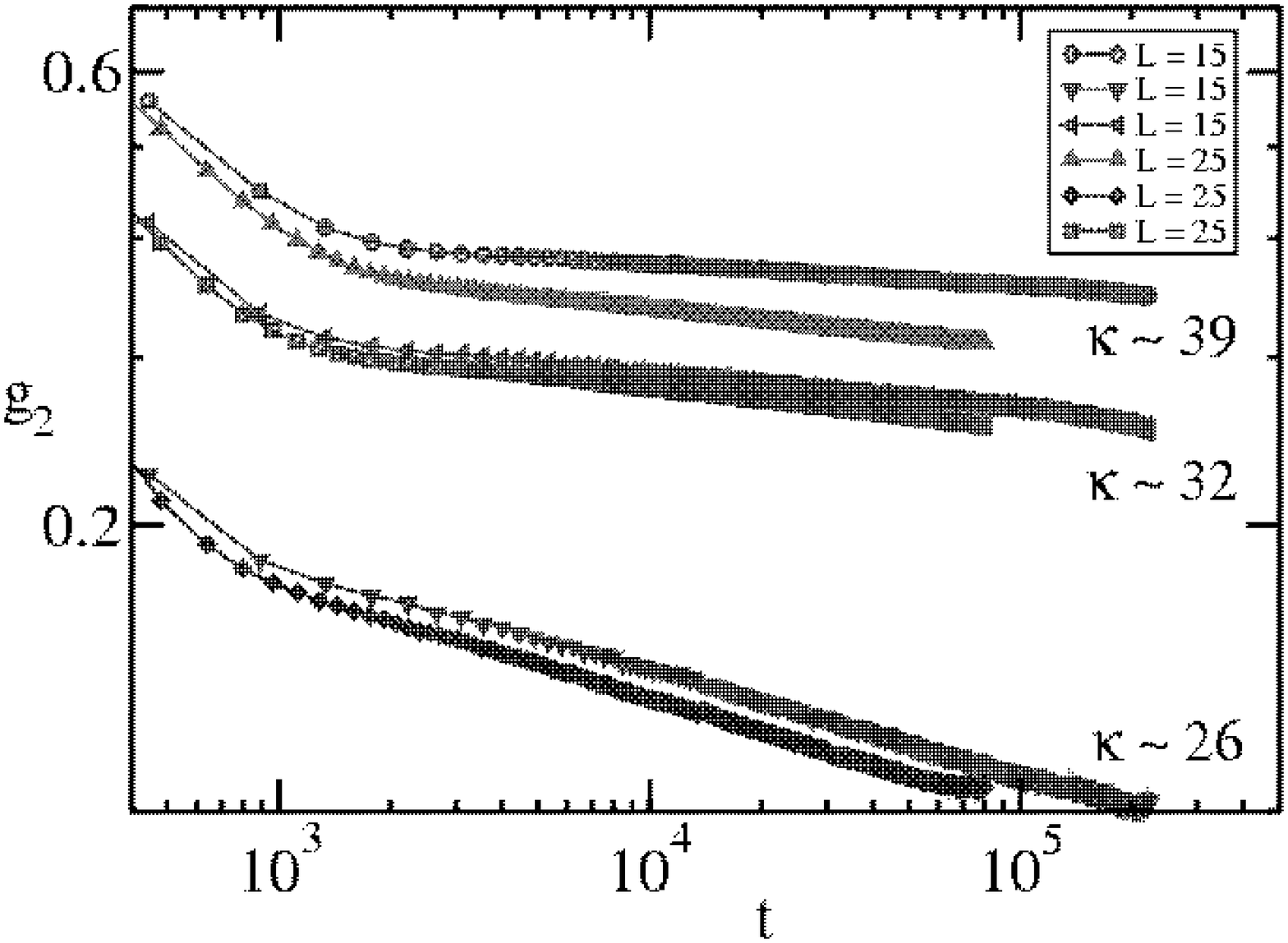}}
\caption{ \small Log-log plots of (a) $g_{2}(r,0) = \langle cos[2(\theta(r,t) - \theta(0,t))]\rangle$ 
and  (b) $g_{2}(0,t) = \langle cos[2(\theta(r,t) - \theta(r,0))]\rangle$ showing the static 
and temporal behaviour respectively of $g_{2}(r,t)$. Data is shown for systems of sizes 15x15 and 25x25, and different values of $\kappa$. The distance $r$ is in units of rod length and time $t$ is in Monte Carlo time steps.}
\label{fig:g2}
\end{figure*}

\begin{figure*}[htbp]
\subfigure[]{
\label{fig:subfig:g4r}
\includegraphics[height=5cm, width=7cm]{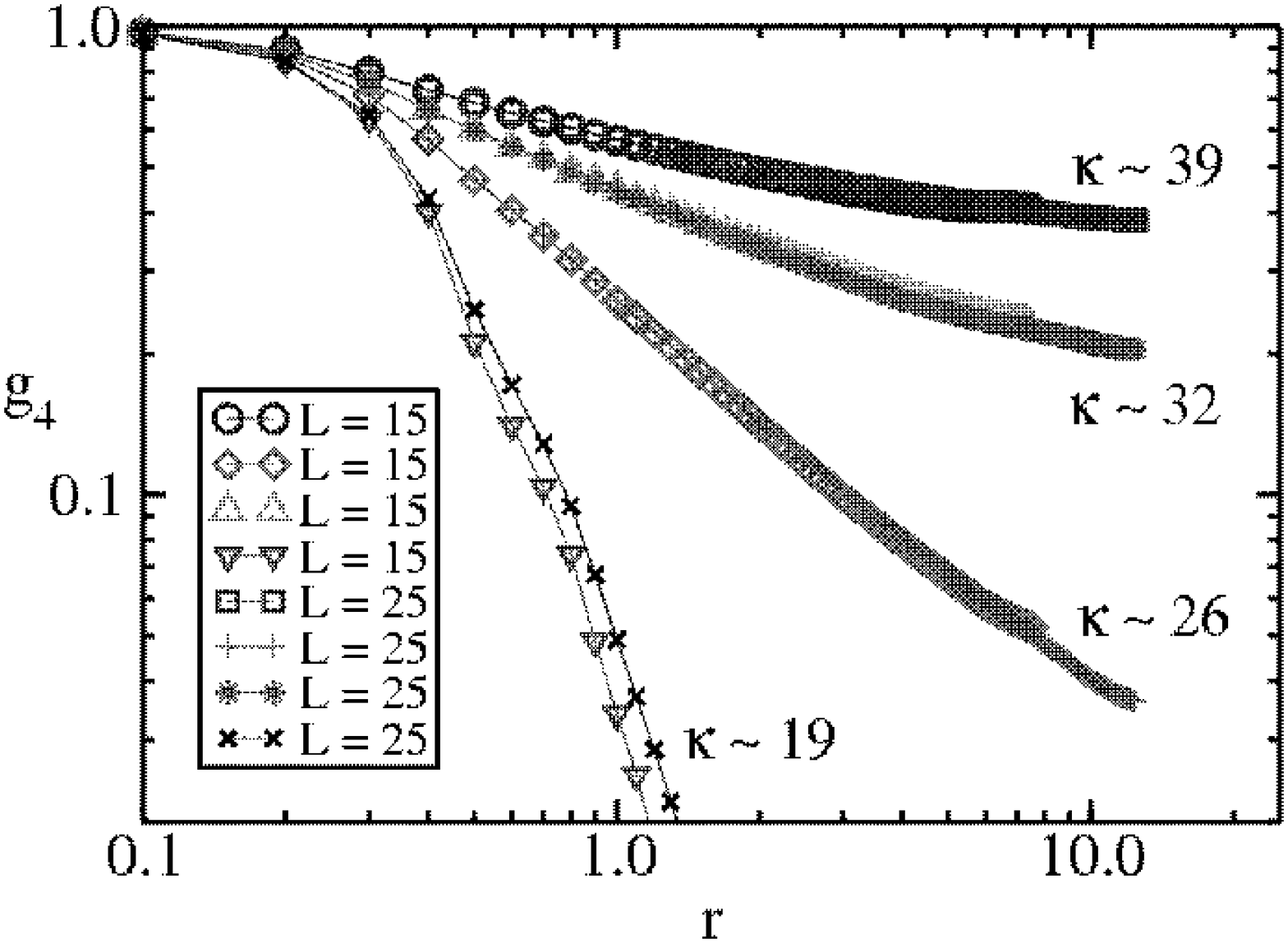}}
\subfigure[]{
\label{fig:subfig:g4t}
\includegraphics[height=5cm, width=7cm]{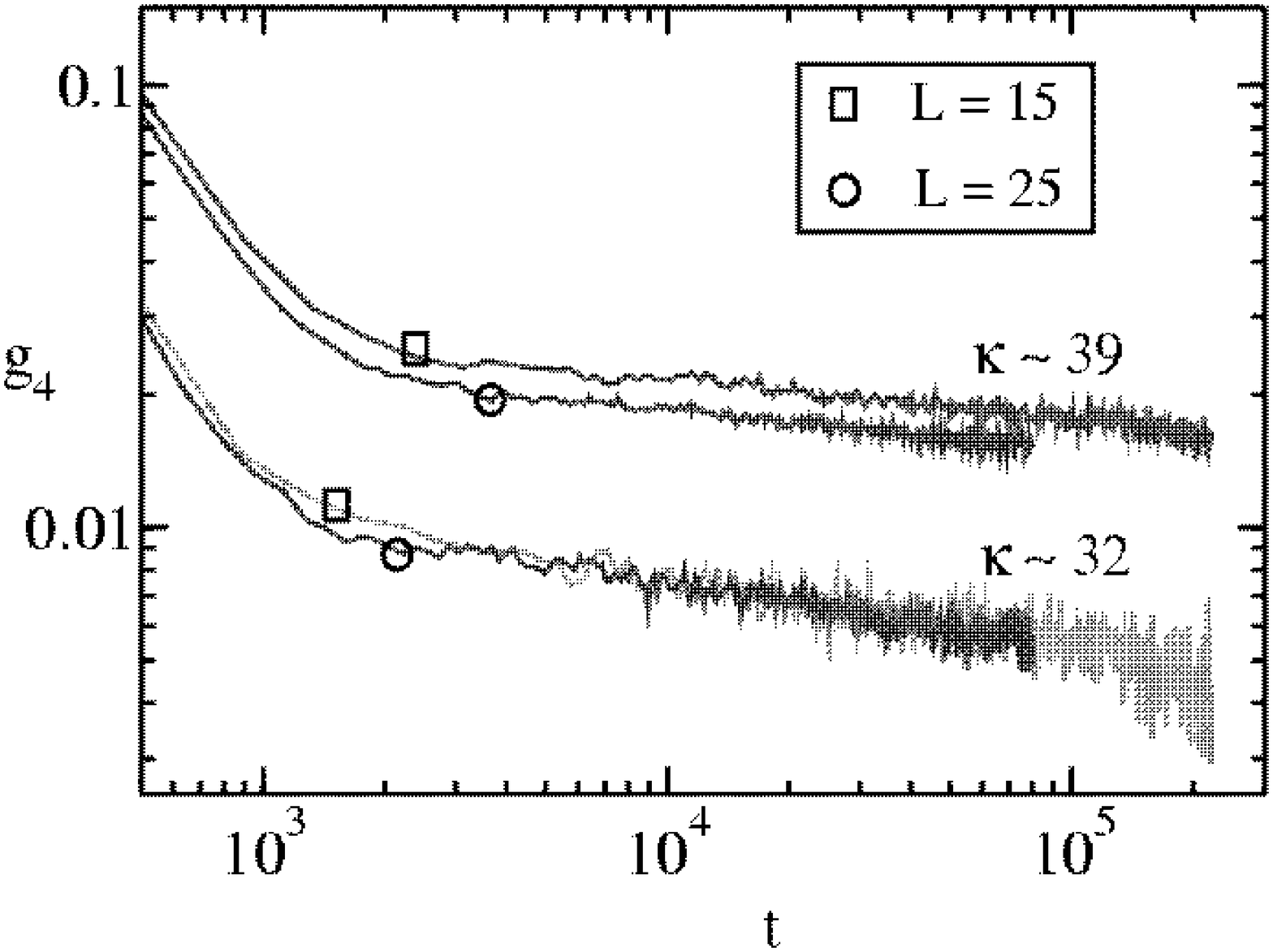}}
\caption{ \small Log-log plots of (a) $g_{4}(r,0) = \langle cos[4(\theta(r,t) - \theta(0,t))]\rangle$ 
and (b) $g_{4}(0,t) = \langle cos[4(\theta(r,t) - \theta(r,0))]\rangle$ showing the static 
and temporal behaviour respectively of $g_{4}(r,t)$, for the same system sizes and values of
$\kappa$ as in Fig. \ref{fig:g2}.} 
\label{fig:g4}
\end{figure*}

  Let us define a general orientational correlation function 
 $g_\ell({\bf r},t) = \langle cos[\ell(\theta({\bf r},t) - \theta(0,0))]\rangle$ where $\ell$ is an even integer.
We studied static and dynamical properties by investigating $g_{\ell}(r,0)$ and $g_{\ell}(0,t)$.
We calculated spatial correlations by forming circular bins around each rod in turn, computing
$g_{\ell}(r,0)$ for each bin, repeating this process for all rods in the configuration and averaging over all rods
(see Fig. \ref{fig:g2}). 
 The dynamical correlation function,  $g_{\ell}(0,t)$ 
was calculated by coarse-graining. The system was divided into a number of small cells (1$\times$1) and an average value of
 orientation was assigned to each cell by averaging over the orientations of those needles whose centres of mass
lie in the cell. The value of $g_{\ell}(0,t)$
was computed using this average value over each time frame, and averaging over all the cells (see Fig. \ref{fig:g4}). 
The initial drops of the curves in Figs. \ref{fig:subfig:g2t} and \ref{fig:subfig:g4t} are sensitive to the size of the cell used, while the power laws
seen at the larger times do not depend on the cell size.

In the nematic-like phase, i.e. for $\kappa$ beyond $\kappa_{c}$, the correlations decay algebraically
$g_\ell(r,0) \sim r^{-\eta_\ell}$ and $ g_\ell(0,t) \sim t^{-\beta_\ell}$. There are pronounced finite size
effects which lead to a flattening of the curves for $r \approx L/2$,  limiting the range over which 
the power law behaviour extends.
We found that the values of the exponents $\eta_{\ell}$ and $\beta_{\ell}$  vary continuously with
 $\kappa$ as shown in Table I. 
The estimations were done over 10 independent sets of configurations, each averaged over 1000 configurations. 
For $\kappa \simeq \kappa_{c} $, we observed that $\eta_{2} \simeq 0.23\pm 0.03 $ close to the predicted
 Kosterlitz-Thouless value 0.25.
Our results for the static case, agree with those reported by Frenkel and Eppenga ~\cite{eppen} for the case of a
fixed number of hard rods on 2D plane. We confirm that at the critical point $\kappa = \kappa_{c}$,
the mean density is $\simeq$ 7~\cite{eppen}.

\begin{table*}
\caption {$\kappa$ dependence of exponents $\eta_{\ell}$ and $\beta_{\ell}$ for $\ell$ = 2 and 4 . 
The estimated error is indicated in the brackets.}   
\begin{ruledtabular}
\begin{tabular}{c  c  c  c  c  c  c}
$\kappa$ & $\rho$ & $\eta_{2}$ & $\eta_{4}$ & $\beta_{2}$ & $\beta_{4}$ & $K/B$                    \\ \\
\hline
26.0  &     7.3   &    0.23 (0.03)  &    0.87 (0.06)   &   0.11 (0.04)  &  0.31 (0.06) &  5.3 $\times 10^{-3}$ \\ \\

32.3  &     9.4   &    0.098 (0.006)  &    0.41 (0.02)   &   0.037 (0.007)  &  0.16 (0.04) &  12.8 $\times 10^{-3}$\\ \\

39.0  &    11.4   &    0.059 (0.003)  &    0.25 (0.01)   &   0.022 (0.003)  &  0.09 (0.01)  & 21.4 $\times 10^{-3}$\\ \\
\end{tabular}
\end{ruledtabular}
\end{table*}
 
For $\kappa > \kappa_{c}$, it was observed that exponents obtained from static
correlations $g_{2}(r)$ and $g_{4}(r)$ are related through $\eta_{2} \simeq \eta_{4}/4$. It was also found that the exponents
derived from the temporal correlations $g_{2}(t)$ and $g_{4}(t)$ are related in a similar way, i.e. $\beta_{2} \simeq
\beta_{4}/4$.  Further, the ratios $\eta_{\ell}/\beta_{\ell}  \simeq $ 2.0 for $\ell$ = 2 and 4, implying that the
dynamical exponent $z_{dyn}$ is 2. These observations can be understood on the basis of the simple model discussed below.

In order to model the dynamics we note that the stochastic evaporation and deposition events change the local value of the  coarse-grained angle field $\theta$
in a noisy, diffusive way. In the discussion below, we take the angle to be an unconstrained variable running from 
-$\infty$ to +$\infty$ with ($\theta + n\pi$) denoting the same needle orientation for integer $n$.
   We consider a simple phenomenological equation
\begin{equation}\label{eq:eqew}
{\partial \theta \over \partial t}  = K {\nabla^{2}\theta}  + \xi    
\end{equation}

\noindent where $\xi$ denotes white noise which satisfies
\begin{equation}\label{eq:eqno}
{\langle \xi({\bf r},t) \xi({\bf r'},t') \rangle} = B {\delta({\bf r-r'}) \delta(t-t')}  
\end{equation}

\noindent where $B$ is a constant.
This is of the same form as the Edwards-Wilkinson equation~\cite{edwar}, which describes the evolution of 
a fluctuating interface. In our context, Eq.(\ref{eq:eqew}) follows from the symmetric form of the Frank free energy
$F$ = $1\over 2$ $K \int (\nabla \theta)^{2} d\theta$ on using the phenomenological Langevin equation
$\partial \theta \over \partial t$ = $-\delta F \over \delta \theta$ + $\xi$.
  A more complete description would involve coupled equations
for the non-conserved density and orientational fields. From Eq.(\ref{eq:eqew}) and (\ref{eq:eqno}) 
it follows that 
\begin{equation}\label{eq:eqrt}
{\langle \theta({\bf r+r'},t+t')\theta({\bf r},t)\rangle } =  {{B \over 64 \pi^{5} K} \int {d^{2}k \over k^{2}}
e^{\it i {\bf k}\cdot {\bf r'}} e^{-Kk^{2} t'}}                          
\end{equation}

\noindent Setting $t'$ = 0 we find that 
\begin{equation}\label{eq:eqrr}
{\langle [\theta({\bf r+r'},t)-\theta(r,t)]^{2} \rangle}  =  {B \over 32 \pi^{5} K} \times {2\pi \ln(r)}   
\end{equation}

\noindent which, using the Gaussian property of $\theta$, further implies
\begin{equation}\label{eq:eqcr}
{\langle cos\{\ell(\theta({\bf r+r'},t)-\theta(r,t))\} \rangle} \simeq {r^{-\eta_{\ell}} }
\end{equation}

\noindent where $\eta_{\ell} $ = $\ell^{2} B$/32$\pi^{4}$$K$. The measured values of $\eta_{\ell}$ can
be used to find $K/B$, whose value is included in Table I.  

Similarly, setting $r'$ = 0 in Eq.(\ref{eq:eqrt}) we find the autocorrelation function
\begin{equation}\label{eq:eqtt}
{\langle cos\{\ell(\theta({\bf r},t+t')-\theta(r,t))\} \rangle} \simeq {t^{-\beta_{\ell}}}    
\end{equation}

\noindent where $\beta_{\ell} $ = $\ell^{2} B$/64$\pi^{4}$$K$.  
Thus, for all $\kappa  > \kappa_{c}$ the ratio of $\eta_{4}$ to $\eta_{2}$ (and $\beta_{4} $
to $\beta_{2}$) is expected to be 4; as we have seen above, our numerical results confirm this. Also, we find
the dynamical exponent $z_{dyn}$ = $\eta_{\ell} /\beta_{\ell} $ is $\simeq$ 2.0 .  

\section{\label{sec:level1} Effect of inhomogeneous $\kappa$}

 In this section, we explore the effects of having a spatial variation 
of the deposition-evaporation rate ratio $\kappa$. In a physical system, such a variation could 
arise from the variation of the chemical potential or substrate temperature from one spot to 
another as their local values could influence the local detachment rate, as seen from Eq. (\ref{eq:eqK}). 
 As expected, such changes in $\kappa$ induce a spatial variation of 
the density; more interestingly, they have a strong effect on the orientational order as well.
We explore these effects by considering several types of spatial variations of $\kappa$.

In parallel to the discussion in section II, let $\Gamma_{d}(x)$ and $\Gamma_{e}(x)$ denote the deposition
 and evaporation rates at point $x$ in the plane, and let $\kappa(x)$ = $\Gamma_{d}(x) \over \Gamma_{e}(x)$. 
Let $C'$ be the configuration reached from configuration $C$ by removing the rod at $x_{m}$ and let $P(C')$ 
and $P(C)$ be the weights of the respective configurations in  steady state. We can check that the condition of 
detailed balance is valid when $P(C)$ has the product form  $P(C)$ = $\Pi_{i} z(x_{i}) $   
where $z(x_{i})$ = $\Lambda^2 \rho(x_{i})\kappa(x_{i})$ is the local fugacity at the location of the $i^{th}$ rod. 
If $C'$ is obtained from $C$ by evaporating the $m^{th}$ rod, then evidently 
$P(C)$ = $P(C') \times z(x_{m})$. 
Recalling that $W(C' \rightarrow C)$ = $\Gamma_{d}(x_{m})$ and $W(C \rightarrow C')$ = $\Gamma_{e}(x_{m})$, we see that 
the condition of detailed balance is valid.

Thus the system reaches an equilibrium steady state which, however, is inhomogeneous in density, due 
to the non-uniform position dependence of $\kappa$. The nature of orientational order depends strongly
on the the way in which $\kappa$ is specified to vary over the plane. Below we consider several 
types of variations 

(i) A single interface separating low and high $\kappa$ regions,
\begin{center}
 $\kappa(x) = \kappa_{1}$, for $x < $$ L \over 2$,

 $\kappa(x) = \kappa_{2}$, for $x > $$ L \over 2$. 
\end{center}

(ii) A uniform gradient in $\kappa$ across the substrate,

\begin{center}
 $\kappa(x) = \kappa_{1}$ (1 + $\alpha$ $x \over L$)
\end{center}

(iii) Random variation of $\kappa$ in the X-direction only, 

\begin{center}
$\kappa(x) = \kappa_{1} + \delta \kappa(x)$
\end{center}

\noindent where $\delta \kappa(x) < \kappa_{1}$ is a random function of $x$.  

(iv) A random binary assignment of $\kappa$ on a grid on the 2D substrate

 In the first three cases, periodic boundary conditions were applied in Y-direction and open boundary 
conditions along X. In the last case, open boundaries were used in both directions. In all the 
cases considered, the ranges of $\kappa$ values were chosen to be above $\kappa_{c}$, the critical value in
the uniform case. Our findings are as follows :

\subsection{\label{sec:level2}  $\kappa_{1} - \kappa_{2}$ interface} 
\begin{figure}[htbp]
\subfigure[]{
\label{fig:subfig:k1-k2-a}
\includegraphics[height=3.7cm, width=6.2cm]{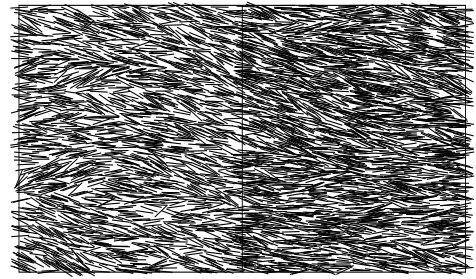}}
\subfigure[]{
\label{fig:subfig:k1-k2-b}
\includegraphics[height=6cm, width=9cm]{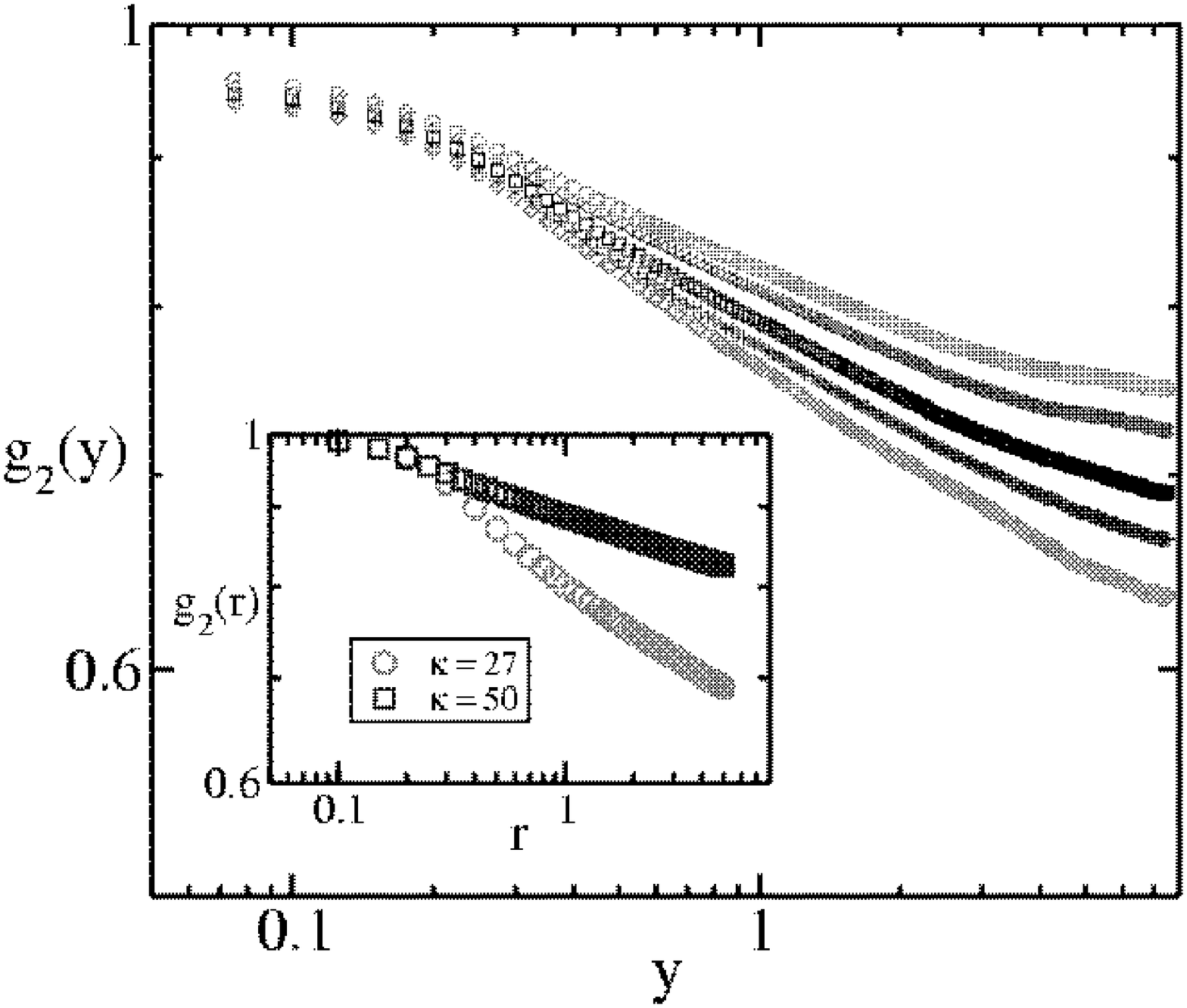}}
\caption{ \small (a) Snapshot of a typical hard rod configuration with a single $\kappa_{1}-\kappa_{2}$ interface.
The system size is 25 $\times$ 15 and $\kappa_{1}$ = 30 (left half) and $\kappa_{2}$ = 50 (right half). Boundary
conditions : Open (along X) , Periodic (along Y). Notice the difference in density in the two halves.
(b)This plot shows the decay of the orientational correlation function 
$ g_{2}({\bf y})$ = $\langle cos[2(\theta({\bf y}) - \theta(0))]\rangle $ calculated for a pair of points in the same 
vertical strip of unit width. Curves from bottom to top correspond to different strips in two halves in 
the configuration (a).
 The inset shows $ g_{2}({\bf r})$ = $\langle cos[2(\theta({\bf r}) - \theta(0))]\rangle $
measured radially for a box of size 6 $\times$ 6 which is positioned at the centre of each of the left and right
halves of the same configuration. The distances $r$ and $y$ are in units of rod length. }
\label{fig:k1-k2}
\end{figure}

 Here, a uniform value $\kappa_{1}$ operates upto half-way across the 2D plane along the X-direction, 
while $\kappa = \kappa_{2} (> \kappa_{1})$ in the remaining half. 
In the vicinity of the interfacial region, the rods are observed to orient in the direction
the of $\kappa$ gradient i.e. perpendicular to the interface (see Fig. \ref{fig:subfig:k1-k2-a}). 

This can be understood on entropic grounds. That arrangement of rods is favoured which maximizes the entropy. 
By symmetry, the preferred average orientation of rods should be either (a) parallel or (b) perpendicular
 to the interface. Consider those rods in the high-$\kappa$ half, whose centres lie very close to the interface
(within half a rod length) so that part of such rods can reach into the low density side.
 Small variations in the angle of each rod would contribute to the entropy, but these are
 limited by the presence of other rods. A horizontal average alignment allows the rods to sample
a less dense environment, and thus be subject to fewer constraints, on one side. Thus, option (b) 
would be preferred over (a).
The effect of interface-induced horizontal alignment is felt for some distance away from the interface on both sides.
This is evident in Fig.\ref{fig:subfig:k1-k2-a}, which shows a steady state configuration in a system of size 25 $\times$ 15 (in units of
rod length), with $\kappa_{1}$ = 30 and $\kappa_{2}$ = 50. However, for a large enough size, the system reverts to
a power-law phase in the region far from the interface as observed in the uniform $\kappa$ case. The 
correlation function decays as a power law in the bulk, away from the interface, as shown in 
Fig.\ref{fig:subfig:k1-k2-b}. 

\begin{figure}[htbp]
\includegraphics[height=3.7cm, width=6.2cm]{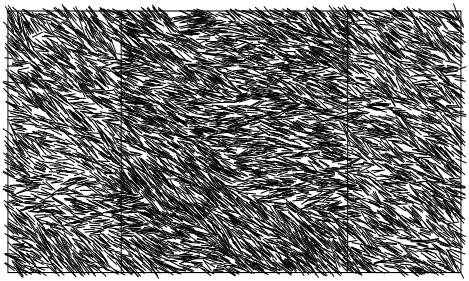}
\caption{\small A dual-interface configuration in an 80 $\times$ 15-sized system with periodic boundary conditions in both 
directions. The middle half of size 40 $\times$ 15 and with $\kappa_2$ = 50 separates two quarters, linked at the
boundary, each of size 20 $\times$ 15 and having $\kappa_1$ = 30.} 
\label{fig:k1-k2-k1}
\end{figure}

\begin{figure}[htbp]
\includegraphics[height=6cm, width=8cm]{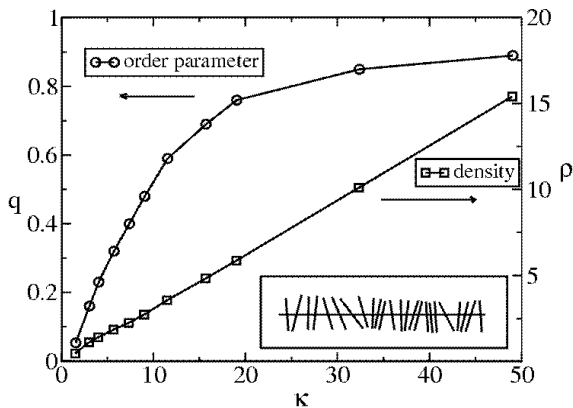}
\caption {\small  Variation of the density($\rho$) and order parameter($q$) with $\kappa$ for the 1D model which 
a limiting case of the 2-D $\kappa_1 - \kappa_2$ model for Fig. \ref{fig:k1-k2-k1}. 
As shown schematically in the inset, the preferred orientation of needles is perpendicular to the
line.}
\label{fig:1d}
\end{figure}

We also considered the case with two values of $\kappa$, using  periodic boundary conditions in both 
directions which leads to two
interfaces (Fig. \ref{fig:k1-k2-k1}). The figure shows a 80 $\times$ 15-sized system with periodically linked 
left and right quarters with $\kappa_1$ = 30, while the middle half has $\kappa_2$ = 50.
Evidently, this system too shows interface-induced horizontal alignment. 
This geometry admits of an interesting limit where the entropy driven alignment is particularly clear. On
shrinking the width of the central region to zero, at the same time taking the limit $\kappa_1 \rightarrow 0$,
we obtain a 1D model as a limiting case. In this model, deposition and evaporation moves are allowed with
needle centres constrained to lie on the line. Needles are found to orient preferentially perpendicular to the
line (see the inset in Fig. \ref{fig:1d}).
The reason is evident. If the mean
orientation of the director is perpendicular to the substrate, needles have the largest leeway to make angular
excursions about the mean - i.e. the rotational entropy is then the largest.
 The variation with $\kappa$ of the density and 
order parameter $q$ = $\langle cos 2\phi \rangle$ where $\phi$ is the angle made with the
direction perpendicular to the line, is shown in Fig. \ref{fig:1d}. 

\begin{figure}[htbp]
\includegraphics[height=3.7cm, width=6.2cm]{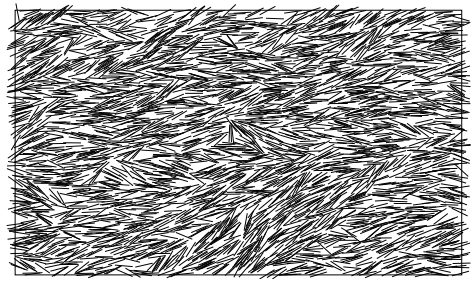}
\caption {\small A typical configuration of size 25 $\times$ 15 with uniform $\kappa = 27$ and open
 boundary conditions along the X-direction. Free boundaries induce alignment which propagates some distance 
into the bulk. }
\label{fig:k-opbc}
\end{figure}

A similar effect should also lead to rods getting aligned horizontally if they are close to an open boundary in the 
2D system.
That this is so can be seen in Fig. \ref{fig:k-opbc}, which shows a system with uniform $\kappa = 27$ and 
open boundary conditions along the X-direction.

\subsection{\label{sec:level2} Uniform $\kappa$ gradient} 
\begin{figure}[htbp]
\includegraphics[height=3.7cm, width=6.2cm]{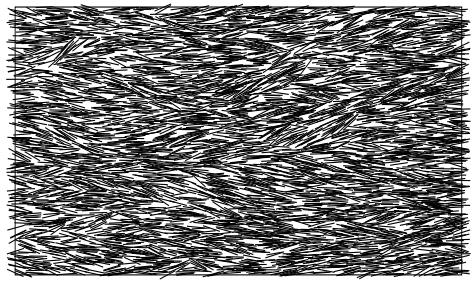}
\caption{\small A typical configuration for a 25 $\times$ 15-sized system with a 
 uniform $\kappa$ gradient, with $\kappa_{L}$ = 32 and $\kappa_{R}$ = 50 at the two ends 
respectively. The horizontal alignment induced by the gradient is evident.}
\label{fig:k-u-grad}
\end{figure}

 This case is related to the discussion above, 
as the linear increase in $\kappa$ may be viewed as a continuous succession of interfaces from one end to the other.
Since each interface induces an alignment of rods across it, this results in overall alignment of the rods 
in the system (see Fig. \ref{fig:k-u-grad}). Note that the alignment is not an outcome of spontaneous breaking of
orientational symmetry, as the gradient in $\kappa$ singles out a direction in space. We checked that the 
horizontal alignment is not tied to the aspect ratio of the container by simulating a system size 10 $\times$ 40,
and observing overall horizontal alignment of rods in the steady state.

\begin{figure}[htbp]
\includegraphics[height=6cm, width=8cm]{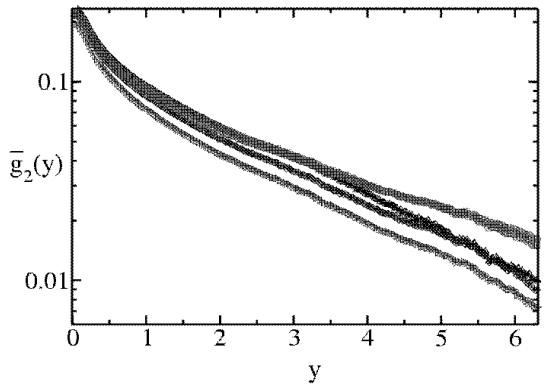}
\caption{ \small  Correlation function in a 25 $\times$ 15-sized system with a uniform $\kappa$ gradient. Shown is
the log-normal plot $\bar{g}_{2}(y) = \langle cos[2(\theta(y) - \theta(0))]\rangle - q_{0}^{2}$ 
for pairs of points in the same vertical strip of unit width. Curves from bottom to top correspond to
 different vertical strips in the order of increasing $\kappa$. 
It is evident from the plots that $\bar{g}_{2}(y)$ exhibits exponential decay. The distance $y$ is in units of rod length.}
\label{fig:k-u-g2r}
\end{figure}

Figure \ref{fig:k-u-grad} shows a steady state configuration in a system of size 25 $\times$ 15, with $\kappa$ varying linearly
from a value $\kappa_{L}$ = 32 at the left end to a value $\kappa_{R}$ = 50 at the right end. In our simulations, 
the system was equilibrated for $10^{7}$ MCS and $10^{4}$ post-equilibration configurations
were used to calculate averages. We studied the spatial and dynamical behaviour of the orientational correlation
function  $g_{2}(r,t) = \langle cos[2(\theta(r,t) - \theta(0,0))]\rangle$.
Since the system is inhomogeneous along the X-direction, the substrate was divided into vertical strips, each having 
width of a rod length, and each strip was studied separately. The Y-density of needles inside each strip was 
uniform though the density varies from strip to strip. We monitored the correlation function 
 $g_{2}(y) = \langle cos[2(\theta(y+y_{0}) - \theta(y_{0}))]\rangle$ (see Fig. \ref{fig:k-u-g2r}), and found that $g_{2}(y)$ decays
exponentially to a non-vanishing constant value $q_{0}$ which differs from strip to strip. For the system under
study, $q_{0}^{2}$ varies in the range 0.68-0.77 over the strips.

 The dynamical correlation function, i.e.  $g_{2}(t) = \langle cos[2(\theta(t) - \theta(0))]\rangle$ 
was calculated by coarse-graining. The system was divided into number of small cells (2$\times$2) and an average value of
 orientation was assigned to each cell by averaging over needles in it. The plot of $g_{2}(t)$ 
is shown in Fig. \label{fig:k-u-g2t}for cells at different values of X. Each curve shows an exponential approach to a non-zero constant value.
The behaviour of both the spatial and dynamical orientational parts of the correlation functions indicates 
a phase with overall orientational alignment. 

\begin{figure}[htbp]
\includegraphics[height=6cm, width=8cm]{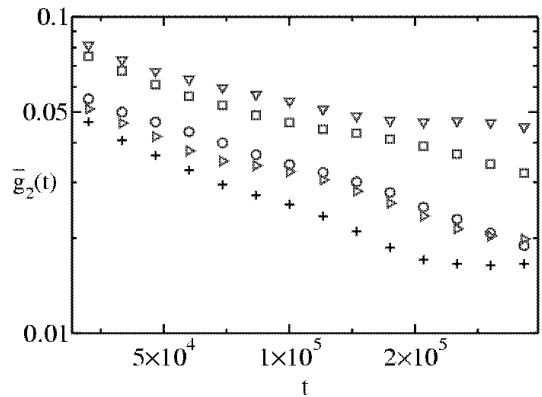}
\caption{ \small Dynamical correlation function in the system of Fig. \ref{fig:k-u-grad} and \ref{fig:k-u-g2r}.
 Log-normal plots of $\bar{g}_{2}(t) = \langle cos[2(\theta(t) - \theta(0))]\rangle - q_{0}^{2}$
as shown, with curves from bottom to top correspond to different cells with increasing $\kappa$
(refer to the main text for details). The time $t$ is in Monte Carlo time steps.}
\label{fig:k-u-g2t}
\end{figure}
 
\subsection{\label{sec:level2} Random variation of $\kappa(x)$ } 

We have seen that a uniform gradient in $\kappa$ results in an orientationally ordered state. However,
the argument for ordering does not depend on the gradient being constant in magnitude or sign. Thus,
if $\kappa$ varies randomly (with $\kappa > \kappa_{c}$) along the X-direction but is uniform along Y, 
the resulting state once again should exhibit horizontal alignment with needles aligned along the X-direction.
The resulting state can once again be viewed as continuous succession of interfaces and 
should display overall alignment.

\begin{figure}[htbp]
\includegraphics[height=3.7cm, width=6.2cm]{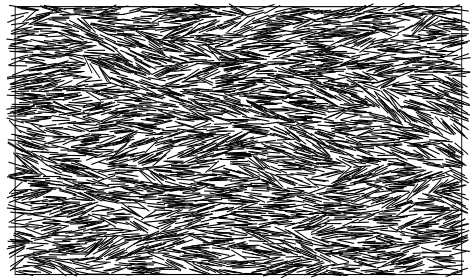}
\caption{\small A typical configuration for a 25 $\times$ 15-sized system with a 
 $\kappa$ gradient achieved by varying the value of $\kappa$  randomly around 32 in the X-direction only. 
 The horizontal alignment induced by the gradient is evident.}
\label{fig:k-r-grad}
\end{figure}

\begin{figure}[htbp]
\includegraphics[height=6cm, width=8cm]{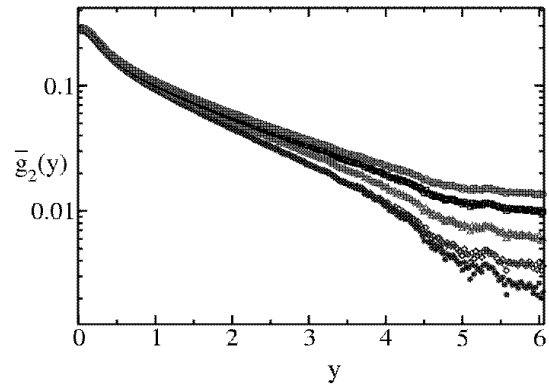}
\caption{ \small  Correlation function in a 25 $\times$ 15-sized system with a random $\kappa$ gradient in the X-direction only.
 Shown is the log-normal plot $\bar{g}_{2}(y) = \langle cos[2(\theta(y) - \theta(0))]\rangle - q_{0}^{2}$ 
for pairs of points in the same vertical strip of unit width. Curves from bottom to top correspond to different 
vertical strips in the order of increasing $\kappa$.
It is evident from the plots that $\bar{g}_{2}(y)$ exhibits exponential decay. The distance $y$ is in units of rod length.} 
\label{fig:k-r-g2r}
\end{figure}

\begin{figure}[htbp]
\includegraphics[height=6cm, width=8cm]{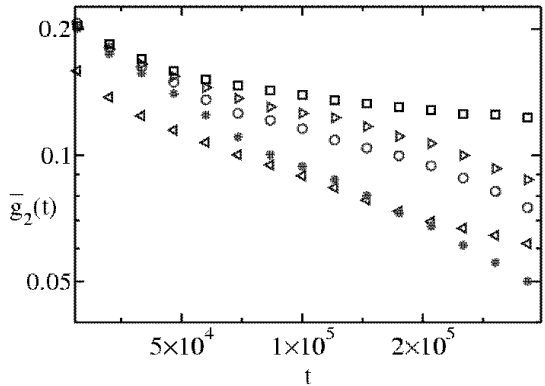}
\caption{ \small Dynamical correlation function in the system of Fig. \ref{fig:k-r-grad} and \ref{fig:k-r-g2r} .
 Log-normal plots of $\bar{g}_{2}(t) = \langle cos[2(\theta(t) - \theta(0))]\rangle - q_{0}^{2}$
as shown, with curves from bottom to top correspond to different cells with increasing $\kappa$. The time $t$ is in
Monte Carlo time steps.}
\label{fig:k-r-g2t}
\end{figure}

 Figure \ref{fig:k-r-grad} shows a steady state configuration obtained with a quenched random variation of $\kappa(x)$.
The system was simulated by varying $\kappa$ randomly around a value of $\kappa$ 
 such that $\kappa \pm \delta \kappa(x) > \kappa_{c}$,
where $\delta \kappa(x)$ denotes random variations along X-direction.  We used $\kappa \simeq 32$ and 
$\delta \kappa = 2.0$. The correlation functions $g_{2}(y)$ and $g_{2}(t)$
behave similarly to the uniform gradient case (see Fig. \ref{fig:k-r-g2r} and\ref{fig:k-r-g2t} ).
 Thus, this case also yields a phase with overall orientational alignment.

\subsection{\label{sec:level2} 2D random binary distribution of $\kappa$ values}
\begin{figure}[htbp]
\subfigure[]{
\label{fig:subfig:k-glass-a}
\includegraphics[height=5cm, width=5cm]{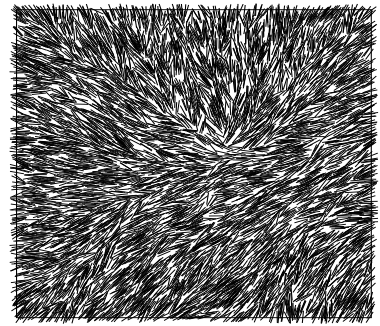}}
\subfigure[]{
\label{fig:subfig:k-glass-b}
\includegraphics[height=5cm, width=5cm]{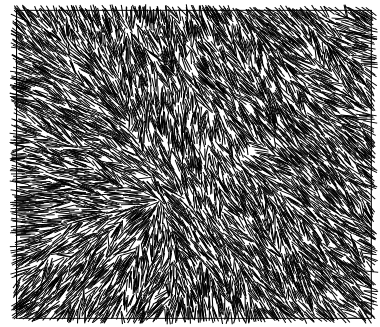}}
\caption {\small Snapshots of hard rod configurations for random binary distribution of $\kappa$ values 27 and 50
on the substrate of size 25 $\times$ 25. Representative configurations characterizing two different states ((a) and (b))
 which are reached from different initial conditions, for the same $\kappa$ distribution.}
\label{fig:k-glass}
\end{figure}

 In this case, the fugacity is set inhomogeneously in a quenched disordered fashion, so that the tendency to 
align locally along gradients results in competing patterns of order, i.e the system is frustrated.
 The resulting state has glassy features and contains domains of different orientations (see Fig. \ref{fig:k-glass}) 
~\cite{binde4}. 

\begin{figure}[htbp]
\includegraphics[height=6cm, width=8cm]{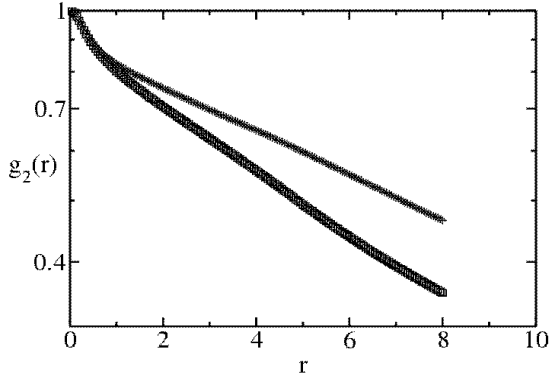}
\caption{ \small Evidence for exponential decay of spatial correlations in a system with random binary distribution of $\kappa$.
The figure shows log-normal plots of $g_{2}({\bf r}) = \langle cos[2(\theta({\bf r}) - \theta(0))]\rangle$.
The parameters are the same as in Fig. \ref{fig:k-glass}, and the curves correspond to the different steady states evolved
from two different initial conditions. The distance $r$ is in units of rod length.}
\label{fig:k-glass-g2r}
\end{figure}

\begin{figure}[htbp]
\includegraphics[height=6cm, width=8cm]{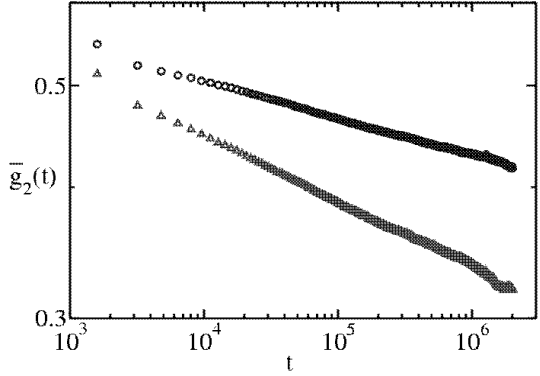}
\caption{ \small Evidence for a power law decay of temporal correlations in a system with random binary distribution of $\kappa$.
The figure shows log-log plots of $\bar{g}_{2}(t) = \langle cos[2(\theta(t) - \theta(0))]\rangle - q_{0}^{2}$.
The parameters are the same as in Fig. \ref{fig:k-glass}, and the curves are obtained in the different steady states 
reached from two different initial conditions. The time $t$ is in Monte Carlo time steps.}
\label{fig:k-glass-g2t}
\end{figure}

In our simulations, we divided the substrate of size 25 $\times$ 25 into a grid of unit length squares.
Each square was randomly assigned a $\kappa$ value either 27 or 50 (both greater than $\kappa_{c}$). 
  The random $\kappa$ gradient across square edges generates local disorder, which
can disrupt the orientational order and result in
destruction of orientational alignment on the scale of the system size. 
The effect of quenched random disorder due to orientational
 randomness of cross-links in a system of nematic elastomers has been studied earlier~\cite{y-kyu}  and the model was
reported to have spin-glass like behaviour. In our model, the disorder emerges from the randomness in
the spatial distribution of $\kappa$ values. We find that
the spatial correlation $g_{2}(r)$ decays exponentially to zero (Fig. \ref{fig:k-glass-g2r}) whereas the dynamical part $g_{2}(t)$ 
seems to decay in an algebraic manner to a non-zero value (Fig.\ref{fig:k-glass-g2t}).  
                                                                                                                                 
The behavior is suggestive of a glassy system
which is disordered in space but relaxes slowly in time. Moreover, it was also found that with same quenched 
disorder arrangement,
different initial conditions lead to different states. Typical configurations in each of these states are shown
in Fig. \ref{fig:k-glass}, which is a glass-like feature. 

\section{\label{sec:level1} Continuum Description}

As discussed above, our simulations show that a spatially inhomogeneous deposition-evaporation 
ratio $\kappa$ can induce nematic order
and other interesting orientational patterns in the equilibrium state of a system of hard needles. It is
interesting to ask whether these effects can be captured within a phenomenological coarse-grained description
based on including symmetry-allowed terms in the free energy. In the context of liquid crystals, such an approach
has proved successful in studying large-distance phenomena, including the
effects of walls and other inhomogeneities~\cite{gennes}. Below we sketch such a description for our system of 
interest~\cite{yhat}. Besides showing that gradients in the deposition-evaporation rates lead to orientational ordering,
the treatment suggests the occurrence of local splay.

Let us define a nematic director
field $\hat n(r)$ to describe the local coarse-grained value of the orientation of needles (evidently,
$\hat n(r)$ and $-\hat n(r)$ describe the same configuration). In the absence of externally imposed
inhomogeneities, spatial variations of $\hat n(r)$ lead to a free energy described by the Frank
form~\cite{gennes}  

\begin{equation}\label{eq:eqfk}
{ F_{K}} = {\int d^{2}r} [{K_{1} \over 2} {(\nabla \cdot \hat n)^{2}} + {K_{3} \over 2}
               {(\hat n \times (\nabla \times \hat n))^{2}] } 
\end{equation}

\noindent The two terms describe, respectively, contributions of splay and bend to the free energy; there is no
contribution from twist in 2-d. 

 Inhomogeneities in deposition-evaporation rates lead to spatial gradients $\nabla \kappa$, which imply 
new terms in the free energy. These terms consist of scalars involving $\nabla \kappa$ and $\hat n$, 
respecting invariance under $\hat n \leftrightarrow -\hat n$. Two such scalars are obtained by replacing 
the gradient operator by $\nabla \kappa$ in the terms in Eq.(\ref{eq:eqfk}) to get

\begin{equation}\label{eq:eqfj}
{ F_{J}} = {\int d^{2}r} {J_{1} \over 2 }{[(\nabla \kappa \cdot \hat n)^{2}]} +
 {\int d^{2}r}{ J_{3} \over 2}{ [(\hat n \times (\nabla \kappa \times \hat n))^{2}]  } 
\end{equation}

In addition to these terms, which are quadratic in $\nabla \kappa$, one can also construct scalars which involve
$\nabla \kappa$ linearly~\cite{gennes2}  

\begin{equation}\label{eq:eqfl}
{F_{L}} = {\int d^{2}r} {L_{1} \over 2} {\nabla \kappa} {\cdot} {[\hat n (\nabla \cdot \hat n)]} + 
          {\int d^{2}r} {L_{3} \over 2} {\nabla \kappa} {\cdot} {[(\hat n \cdot \nabla) \hat n]}.
\end{equation} 

\noindent Symmetry considerations alone do not suffice to determine the values of the 
coefficients $K_{1}, K_{3}, J_{1}, J_{3}, L_{1}$ and $L_{3}$. 
Their density dependences can be found on noting that 
a change in $\kappa$ induces a change in density, thereby influencing the elastic energy. To incorporate
this, we replace $\hat n$ in Eq. (\ref{eq:eqfk}) by 
($\rho(r) \hat n /\rho_{0}$) where $\rho(r)$ and $\rho_{0}$ are the local and average densities respectively. 
Appendix A contains the resulting form of the free energy $F_{\rho}$ and the values of the coupling constants.

Let us turn to the consequences of the new terms.
With a uniform spatial gradient, $\nabla \kappa$ = $\alpha \hat x$  (case (ii) above), 
$F_{J}$ induces overall alignment of needles. To see this, consider $F_{K} + F_{J}$.
Evidently, $F_{K}$ is minimized by any arrangement in which $\hat n$ is uniform in space while $F_{J}$ is orientation
dependent. 
Writing $\hat n$ = $\hat x cos\phi $ + $\hat y sin\phi$, we find
\begin{equation}\label{eq:eq14}
 {F_{J}} = {\alpha^{2}} ({J_{1} cos^{2}\phi} + {J_{3} sin^{2}\phi }) 
\end{equation}
$F_{K}$ + $F_{J}$ is minimized by having an aligned state, either with $\phi = 0$ (if $J_{3} > J_{1}$) or with
$\phi$ = $\pi \over 2$ (if $J_{3}< J_{1})$. Our numerical results, supported by the entropic considerations 
given above, show alignment in the direction of the gradient, implying $\Delta J$ $\equiv$ $J_{3}-J_{1}$ is positive.

Now consider the effect of $F_{L}$. Setting $\nabla \kappa$ = $\alpha \hat x$ in Eq. (\ref{eq:eqfl}), we find
\begin{eqnarray}
 {F_{L}} = {-\alpha (L_{1} + L_{3})} \quad {sin\phi} \quad {cos\phi} \quad      
{\partial \phi \over \partial x}  \nonumber \\
+ \quad {\alpha (L_{1} cos^{2}\phi - L_{3} sin^{2}\phi)}\quad {\partial \phi \over \partial y}
\end{eqnarray}
$(F_{K}+F_{L})$ can be minimized on noting that each of
the $L_{1}$ and $L_{3}$ terms is an eigenfunction of the Frank elastic matrix. The result is 
\begin{equation}\label{eq:eqpy}
{\partial \phi \over \partial y} = {- L_{1} \over K_{1}}{cos^{2}\phi} + {L_{3} \over K_{3}}{sin^{2}\phi} 
\end{equation}
 
\begin{equation}\label{eq:eqpx}
{\partial \phi \over \partial x} = ({L_{1} \over K_{1}} + {L_{3} \over K_{3}}) \quad {cos\phi \quad sin\phi}
\end{equation}
These terms describe a spiralling tendency of the director in space.

The full problem involves minimizing $F_{J}+F_{K}+F_{L}$.
 If $F_{J}$ is dominant, the director is primarily 
aligned along the gradient implying $\phi$ is small. Eqs.(\ref{eq:eqpy}) and (\ref{eq:eqpx}) then reduce to
$\partial \phi /\partial y \approx -L_{1}/K_{1} $ which describes a spiralling director; 
further, $F_{J}$ restricts angular excursions to be at most $\phi_{0} \propto $$1 \over \sqrt{\Delta J}$.
Thus the predicted state is one with overall alignment along the gradient, but with local splay structures, each with
a small opening angle $\simeq 2\phi_{0}$. This picture is borne out by our simulations.   

 For case (iii)
in which $\kappa(x)$ varies randomly, we see that $F_{J}$ = $\overline{\alpha^{2}}$($J_{1} cos^{2}\phi $ +
 $J_{3} sin^{2}\phi $) where $\overline{\alpha^{2}}$ is the spatial average of the mean squared gradient. As for 
case (ii), the free energy is minimized by having a state with alignment along the gradient, as observed in our
simulations, provided
$J_{3} > J_{1} $.

 In case (iv), the gradients that appear in Eq. (\ref{eq:eqfj}) are random in direction leading to frustration in the 
arrangement of needles. 
Equations (\ref{eq:eqfk}), (\ref{eq:eqfj}) and (\ref{eq:eqfl}) provide a starting point for a theoretical description
of the glassy state that results.

\section{\label{sec:level1} Conclusion}

In summary, we have studied orientational ordering in a 2D grand canonical system of hard rods using deposition
and evaporation moves.  
The control parameter is the ratio $\kappa$ of deposition and evaporation rates, which controls the density.
The system with uniform $\kappa$ displays a transition from an isotropic phase (for $\kappa < \kappa_{c}$) to a phase 
characterized by algebraically decaying  static and dynamical orientational correlations for $\kappa > \kappa_{c}$.
 Further, the values of 
the critical exponents and the behaviour of the orientational cumulant are consistent with Kosterlitz-Thouless theory.
 The numerical results for the dynamical correlation functions are described by a phenomenological
Edwards-Wilkinson equation for the non-conserved orientational field. 
                                                                                                                      
Our principal results pertain to the new behavior induced by having  a position-dependent
$\kappa$, and hence a space-varying density of rods. An anisotropic variation of $\kappa$
(say along the X-direction only) results in needles aligning along the $\kappa$ gradient. 
This was understood by first considering the effect of an interface
separating regions with two values of $\kappa$.  Entropic
considerations lead the needles to align normal to the interface,
i.e. along the gradient.  From another point of view,
$\kappa$-gradients lead to new terms in the Frank-like free energy and
these in turn imply orientational ordering.  Finally in a system with
quenched disorder corresponding to spatially random $\kappa$, we found
indications of orientationally frozen states with glass-like
characteristics.  It would be useful to have a better characterization
and understanding of this glassy state.

The mechanism behind gradient-induced orientational ordering is 
simple: spatial variations of $\kappa$ induce variations in
needle density; and an average alignment of needles along the gradient
is preferred as this leads to an enhanced entropy of rotational
excursions around the mean.  In effect, the $\kappa$-gradient thus
behaves like an external field acting to produce nematic order,
ultimately due to the strong coupling between spatial and
orientational degrees of freedom in the needle system.
Gradient-induced ordering effects should be present in
three-dimensional systems as well.  In $3D$, the Onsager mechanism for
nematic long range order would predict ordering for values of a
uniform $\kappa$ exceeding a critical value $\kappa_c$.  The addition
of a uniform $\kappa$-gradient would be expected to lead to a nonzero
value of nematic ordering for all values of $\kappa$, and to enhance
its value for $\kappa > \kappa_c$.  It would be interesting to test
this prediction, and have a quantitative measure of gradient-induced
ordering in $3D$.

\vspace{1.75cm}
{\bf ACKNOWLEDGMENTS }
\vspace{0.5cm}

 We acknowledge very useful discussions with Yashodhan Hatwalne, Sriram Ramaswamy and Gautam Menon. We thank Deepak
Dhar for a critical reading of the manuscript and suggestions.

\vspace{1.75cm}
{\bf APPENDIX  A}
\vspace{0.5cm}

To incorporate the effect of spatial variation of the density, we write $(\rho(r) \hat n /\rho_{0})$ in place of
the director $\hat n$ in the Frank free energy $F_{K}$ of Eq. (\ref{eq:eqfk}).
The resulting expression $F_{\rho}$ for the free energy can be written as 

\begin{equation}\label{eq:eq16}
{ F_{\rho}} = {\int d^{2}r} [{K_{1} \over {2\rho_{0}^{2}}} {(\nabla \cdot \rho \hat n)^{2}} +
            {K_{3} \over {2\rho_{0}^{4}}} {(\rho \hat n \times (\nabla \times \rho \hat n))^{2}] } 
\end{equation}

The expansion of the integrand involves ten terms :

[$K_{1} \over {2\rho_{0}^{2}}$ $\rho^{2}$ $sin^{2}\phi$ + ($K_{3}\over {2\rho_{0}^{4}}$ $\rho^{2}$) 
              $\rho^{2} cos^{2}\phi$] $(\partial \phi / \partial x)^{2}$ 

+

[$K_{1} \over {2\rho_{0}^{2}}$ $\rho^{2}$ $cos^{2}\phi$ + ($K_{3}\over {2\rho_{0}^{4}}$ $\rho^{2}$) 
              $\rho^{2} sin^{2}\phi$] $(\partial \phi / \partial y)^{2}$ 

+

[$K_{1} \over {2\rho_{0}^{2}}$ $cos^{2}\phi$ + ($K_{3}\over {2\rho_{0}^{4}}$ $\rho^{2}$) 
              $ sin^{2}\phi$] $(\partial \rho / \partial x)^{2}$ 

+

[$K_{1} \over {2\rho_{0}^{2}}$ $sin^{2}\phi$ + ($K_{3}\over {2\rho_{0}^{4}}$ $\rho^{2}$) 
              $ cos^{2}\phi$] $(\partial \rho / \partial y)^{2}$ 

+

[$-K_{1} \over {2\rho_{0}^{2}}$  + ($K_{3}\over {2\rho_{0}^{4}}$ $\rho^{2}$) 
              ] $2\rho$ $sin\phi$ $cos\phi$ $(\partial \phi / \partial x)$ 
               $(\partial \rho / \partial x)$ 

+

[$K_{1} \over {2\rho_{0}^{2}}$  - ($K_{3}\over {2\rho_{0}^{4}}$ $\rho^{2}$) 
              ] $2\rho$ $sin\phi$ $cos\phi$ $(\partial \phi / \partial y)$ 
               $(\partial \rho / \partial y)$ 

+

[$K_{1} \over {2\rho_{0}^{2}}$  - ($K_{3}\over {2\rho_{0}^{4}}$ $\rho^{2}$) 
              ] $2$ $sin\phi$ $cos\phi$ $(\partial \rho / \partial x)$ 
               $(\partial \rho / \partial y)$ 

+

[$-K_{1} \over {2\rho_{0}^{2}}$  + ($K_{3}\over {2\rho_{0}^{4}}$ $\rho^{2}$) 
              ] $2\rho^{2}$ $sin\phi$ $cos\phi$ $(\partial \phi / \partial x)$ 
               $(\partial \phi / \partial y)$ 

+
                                                                                                                           
[$-K_{1} \over {2\rho_{0}^{2}}$ $sin^{2}\phi$  - ($K_{3}\over {2\rho_{0}^{4}}$ $\rho^{2}$) $cos^{2}\phi$
              ] $2\rho$ $(\partial \phi / \partial x)$
               $(\partial \rho / \partial y)$
                                                                                                                           
+
                                                                                                                           
[$K_{1} \over {2\rho_{0}^{2}}$ $cos^{2}\phi$  + ($K_{3}\over {2\rho_{0}^{4}}$ $\rho^{2}$) $sin^{2}\phi$
              ] $2\rho$ $(\partial \rho / \partial x)$
               $(\partial \phi / \partial y)$

In case (ii) where $\kappa$ varies linearly with $x$, the density gradient is non-zero only along the
X-direction, and vanishes along Y, so the terms involving $\partial \rho /\partial y$ do not contribute.
We can then read off the density dependence induced in the elastic constants in terms of the original 
Frank's constants as

$K'_{1}(\rho)$ = $K_{1} \over {\rho_{0}^{2}}$ $\rho^{2}$  and $K'_{3}(\rho)$ = $K_{3}\over {\rho_{0}^{4}}$ $\rho^{4}$
                                                                                                                 
Now, comparing the third term of the expression with Eq. (\ref{eq:eqfj}) and writing $\partial \rho/\partial x$ = $\alpha \zeta(\rho)$ 
where $\zeta(\rho)$ = $\partial \rho / \partial \kappa$, we have
\begin{equation}\label{eq:eq19}
{J_{1}(\rho)} = {K_{1} \over {\rho_{0}^{2}}} {(\zeta(\rho))^{2}} \quad;\quad
{J_{3}(\rho)} = {K_{3}\over {\rho_{0}^{4}}} {\rho^{2}} {(\zeta(\rho))^{2}}
\end{equation}                                                                                                                           
Similarly, grouping the fifth and the tenth terms together and comparing with Eq. (\ref{eq:eqfl}), we obtain
\begin{equation}\label{eq:eq20}
{L_{1}(\rho)} ={K_{1} \over {\rho_{0}^{2}}} {2 \rho} {\zeta(\rho)} \quad;\quad
{L_{3}(\rho)} =(-{K_{3}\over {\rho_{0}^{4}}} {\rho^{2}}) {2 \rho} {\zeta(\rho)}.
\end{equation}


\begin{thebibliography}{99}
\bibitem{ising}  For instance, for Ising-like systems, for $T > T_{c}$ and $L \gg \xi$, $P_{L}(s)$
 approaches a Gaussian distribution, while for $T < T_{c}$ and $L \gg \xi$, 
 $P_{L}(s)$ approaches two Gaussians centered
 at $\pm M_{0}$ where $\xi$ denotes the correlation length and $M_{0}$ is the spontaneous magnetization
 of infinite system.  At criticality, $P_{L}(s)$ approaches a scaled
 universal form for large $L$~\cite{binde3,weber}. Note that $'s'$ denotes the order parameter corresponding to 
 Ising-like systems.
\bibitem{gennes2} The form of these terms is the same as those describing the coupling of a flexoelectric 
nematic to the electric field; see Section 3.3.2 of Ref.~\cite{gennes}. In our case, the external field is
$\nabla \kappa$.
\end{thebibliography}
\end{document}